\documentclass[preprint,aps,amsmath,byrevtex,pra,titlepage,
nofootinbib]{revtex4-1}

\usepackage{dcolumn}
\usepackage{amsmath}
\usepackage{amssymb}
\usepackage{bm}
\usepackage{hyperref}
\usepackage{graphicx}
\usepackage{slashed}

\newcommand{\beq}{\begin{equation}}
\newcommand{\eeq}{\end{equation}}
\newcommand{\eq}[1]{Eq.~(\ref{#1})}

\begin{document}

\title {Three-Loop Contribution to Hyperfine Splitting in Muonium: Polarization Corrections to Light by Light Scattering Blob}
\author {Michael I. Eides}
\altaffiliation[Also at ]{the Petersburg Nuclear Physics Institute,
Gatchina, St.Petersburg 188300, Russia}
\email[Email address: ]{eides@pa.uky.edu, eides@thd.pnpi.spb.ru}
\affiliation{Department of Physics and Astronomy,
University of Kentucky, Lexington, KY 40506, USA}
\author{Valery A. Shelyuto}
\email[Email address: ]{shelyuto@vniim.ru}
\affiliation{D. I.  Mendeleyev Institute for Metrology,
St.Petersburg 190005, Russia}
%\date{}

\begin{abstract}
We calculate corrections  of order $\alpha^3(Z\alpha)E_F$ to hyperfine splitting in muonium generated by the gauge invariant set of diagrams with polarization insertions in the light by light scattering diagrams. This nonrecoil contribution turns out to be $-2.63$ Hz. The total contribution of all known corrections of order $\alpha^3(Z\alpha)E_F$ is equal to $-4.28$ Hz.

\end{abstract}

%\pacs{12.20Ds,31.30.jf,32.10.Fn,36.10.Ee}
%\keywords{hyperfine splitting}

\preprint{UK/12-07}

\maketitle

\section{Introduction}

The hyperfine splitting in muonium is one of the intervals best studied both experimentally and theoretically. Theoretical expression for the hyperfine splitting can be calculated in the QED framework in the form of a perturbation theory expansion in $\alpha$, $Z\alpha$, $m_e/m_\mu$. Current theoretical uncertainty of this expansion is  estimated to be about 70-100 Hz, respective relative error does not exceed $2.3\times10^{-8}$ (see discussions in \cite{egs2001,egs2007,mtn2012}).  The experimental error of the best measurements \cite{mbb,lbdd} of the  muonium HFS is in the interval 16-51 Hz. A new higher accuracy measurement of muonium HFS is now planned at J-PARC, Japan \cite{shimomura}. Combining muonium HFS theory and experiment one can determine the value of $\alpha^2(m_\mu/m_e)$ with the uncertainty that is dominated by $2.3\times 10^{-8}$ relative uncertainty of the HFS theory \cite{mtn2012}. This is currently the best way to determine the precise value of the electron-muon mass ratio. Further reduction of the uncertainty of this mass ratio requires improvement of the HFS splitting theory. Main sources of the theoretical uncertainty are due to still unknown three-loop purely radiative contributions, three-loop radiative-recoil contribution, and nonlogarithmic recoil contributions (see detailed discussion in \cite{egs2007,mtn2012}). We consider reduction of the theoretical error of HFS splitting in muonium to about $10$ Hz as the current goal  of the HFS theory.

\begin{figure}[htb]
\includegraphics
[height=3cm]
{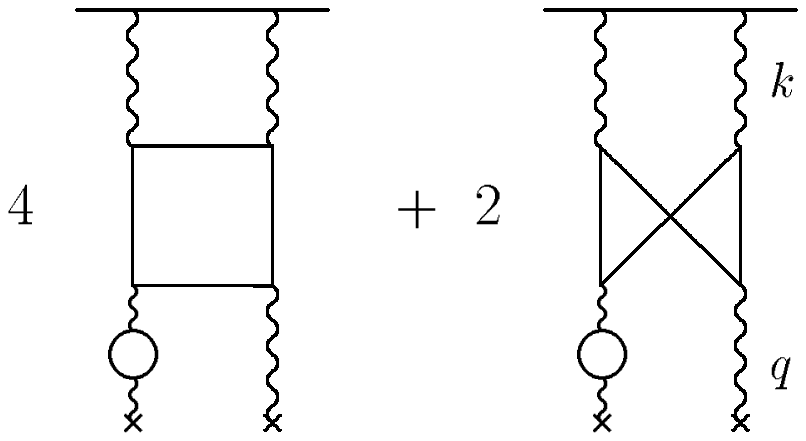}
\caption{\label{downphot}}
\end{figure}

\begin{figure}[htb]
\includegraphics[height=3cm]{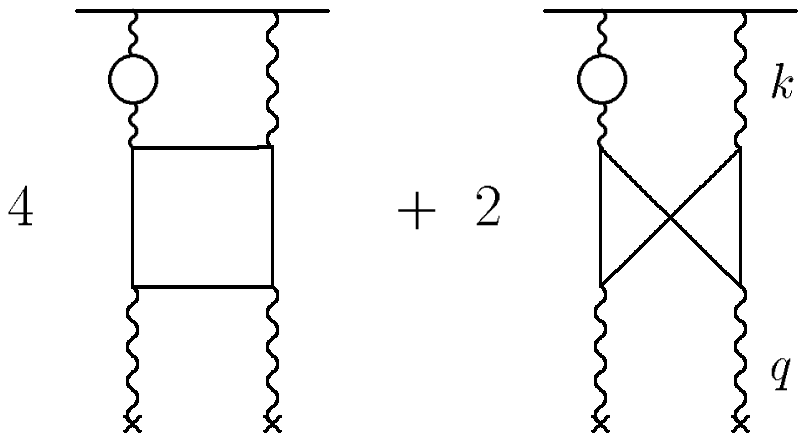}
\caption{\label{upphot}}
\end{figure}

As a step in this direction we calculate below a three-loop contribution to HFS generated by the light by light scattering diagrams in Figs. \ref{downphot} and \ref{upphot} with insertions of one loop polarization in the upper and lower photon lines, respectively.

\section{Calculations}

\begin{figure}[htb]
\includegraphics
[height=3.cm]
{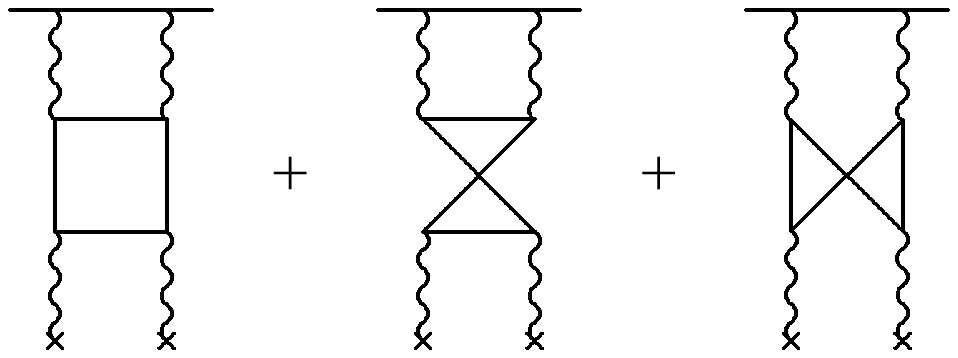}
\caption{\label{lbl}}
\end{figure}

We start with the light by light scattering contribution to HFS that was calculated long time ago \cite{eks1991}. It is generated by the diagrams in Fig. \ref{lbl}, where we have not shown explicitly three more diagrams with the crossed photon lines. In our calculations below we will follow the general approach developed in \cite{eks1991} and start with the light by light scattering contribution  in Fig. \ref{lbl} (see \cite{eks1991})

\beq \label{lblskel}
\Delta E=\frac{\alpha^2(Z\alpha)}{\pi}E_F\frac{3}{64\pi^2}\int \frac{d^4 k}{\pi^2i}\frac{\langle\gamma_\alpha\slashed k \gamma_\beta\rangle}{k^4}
\left(\frac{1}{k^2+2k_0}+\frac{1}{k^2-2k_0}\right)\int \frac{d^3q}{4\pi}\frac{\langle\gamma_\mu\slashed q \gamma_\nu\rangle}{q^4}
S^{\alpha\beta\mu\nu},
\eeq

\noindent
where $k^\mu$ is the four-momentum carried by the upper photon lines, $q^\mu=(0,\bm q)$ is the spacelike four-momentum carried by the lower photon lines, $S^{\alpha\beta\mu\nu}$ is the light by light scattering tensor, and all momenta are measured in the electron mass units. The Fermi energy is defined as

\beq
E_F=\frac{8}{3}(Z\alpha)^4 (1+a_\mu)\frac{m_e}{m_\mu}\left(\frac{m_r}{m_e}\right)^3m_ec^2,
\eeq

\noindent
where $a_\mu$ is the muon anomalous magnetic moment. The angle brackets in \eq{lblskel} denote the projection of the $\gamma$-matrix structures on the HFS interval (difference between the states with the total spin one and zero).

With account for three more diagrams with crossed photon lines not shown explicitly in Fig. \ref{lbl} contributions to HFS of the first two diagrams coincide and we can represent the light by light block as a sum of two contributions corresponding to the first two (ladder) diagrams in  Fig. \ref{lbl} and corresponding to the crossed (last) diagram in Fig. \ref{lbl}

\beq \label{lblintegral}
S^{\alpha\beta\mu\nu}=\int \frac{d^4p}{\pi^2i}\left(2L^{\alpha\beta\mu\nu}+C^{\alpha\beta\mu\nu}\right),
\eeq

\noindent
where (we return to dimensionful momenta here)

\beq
L^{\alpha \beta \mu \nu} = Tr\Biggl[\gamma_{\mu}\frac{1}{{\slashed p}-{\slashed q}-m}\gamma_{\nu}\frac{1}{{\slashed p}-m}\gamma_{\beta} \frac{1}{{\slashed p}-{\slashed k}-m}\gamma_{\alpha}\frac{1}{{\slashed p}-m}\Biggr],
\eeq
\beq
C^{\alpha \beta \mu \nu} = Tr\Biggl[\gamma_{\mu}\frac{1}{{\slashed p}-{\slashed q}-m}\gamma_{\beta}\frac{1}{{\slashed p}-{\slashed q}-{\slashed k}-m}\gamma_{\nu}\frac{1}{{\slashed p}-{\slashed k}-m}\gamma_{\alpha} \frac{1}{{\slashed p}-m}\Biggr].
\eeq

\noindent
Calculating traces we obtain

\beq
\begin{split}
L^{\alpha \beta \mu \nu}& =\left\{ 8D_1^2 g^{\mu \alpha}g^{\nu \beta}
+ 16D_1 g^{\mu \alpha} \left[p^{\nu}q^{\beta} + k^{\nu}p^{\beta}
+ k^{\nu} q^{\beta}-p^{\nu}p^{\beta}\right]
- 8D_1 g^{\mu \alpha}g^{\nu \beta}
\right.
\\
&\times \left[(p\cdot q) + (p\cdot k) + (k\cdot q)\right]+ 32 g^{\mu \alpha}\left[(k\cdot q) p^{\nu}p^{\beta}
- (p\cdot q) k^{\nu}p^{\beta}
-(p\cdot k) p^{\nu} q^{\beta}\right]
\\
&\left.
+ 16 (p\cdot k)(p\cdot q) g^{\mu \alpha}g^{\nu \beta}
- 32 p^{\mu} p^{\alpha} k^{\nu} q^{\beta}\right\}\frac{1}{D_1D_2D_3D_4},
\end{split}
%\]
\eeq

\noindent
and

\beq
\begin{split}
C^{\alpha \beta \mu \nu}
&=\left\{8D_1g^{\mu \alpha}\left[-3k^{\nu} q^{\beta} + k^{\nu} p^{\beta} + p^{\nu} q^{\beta}\right]
- 8D_2  g^{\mu \alpha}k^{\nu}p^{\beta}
- 8D_3  g^{\mu \alpha}p^{\nu}q^{\beta}
\right.
%\eeq
\\
%\[
&\left.
+ 16(k\cdot q)g^{\mu \alpha}p^{\nu}p^{\beta}
- 16(k\cdot q)g^{\mu \alpha}k^{\nu}p^{\beta}
- 16(k\cdot q)g^{\mu \alpha}p^{\nu}q^{\beta}
+ 16p\cdot(k+q)g^{\mu \alpha}k^{\nu}q^{\beta}\right\}
%\]
\\
&\times\frac{1}{D_1D_2D_3D_4},
\end{split}
\eeq

\noindent
where

\beq
D_1 = p^2-m^2,\qquad D_2 = (p-q)^2-m^2,\qquad D_3 = (p-k)^2-m^2,\qquad
D_4 = (p-q-k)^2-m^2.
\eeq

After calculation of the integrals in \eq{lblintegral} we obtain the light by light scattering block in the form

\beq \label{gauginvblb}
S^{\alpha\beta\mu\nu}=2{\cal L}^{\alpha\beta\mu\nu}+{\cal C}^{\alpha\beta\mu\nu},
\eeq

\noindent
where

\beq \label{lblladd}
\begin{split}
{\cal L}^{\alpha\beta\mu\nu}&=\int \frac{d^4p}{\pi^2i}L^{\alpha\beta\mu\nu}
=8\int_0^1 {dx}\int_0^1 {dy}\int_0^1 {dz} \int_0^1 {d\xi}\Biggl\{ - \frac{2y(1-y)}{\Omega(1,y,1,\xi)}(k\cdot q) g^{\mu \alpha}g^{\nu \beta}
%\eeq
\\
%\[
&+\frac{2y^2(1-y)z}{\Omega(1,y,z,\xi)} (k\cdot q)  g^{\mu \alpha}g^{\nu \beta}- \frac{2y(2-y+y^2z)}{\Omega(1,y,z,1)}  g^{\mu \alpha}k^{\nu}q^{\beta} +\frac{y}{\Omega(1,y,z,1)}
%\]
\\
%\[
&
\times\Bigl[k^2 (1-y) + q^2 yz + (k\cdot q)(2-y+yz) \Bigr] g^{\mu \alpha}g^{\nu \beta}
+y^2(1-z)
%\]
\\
%\[
& \times\Biggl[\biggl( -\frac{3}{\Omega(1,y,z,1)} + \frac{2}{\Omega^2(1,y,z,1)}\left[k^2(1-y)^2 + q^2y^2z^2\right]\biggr)
\Bigl[(k\cdot q) g^{\mu \alpha}g^{\nu \beta} - 2 g^{\mu \alpha}k^{\nu}q^{\beta} \Bigr]
%\]
\\
%\[
&+\frac{2(1-y)yz}{\Omega^2(1,y,z,1)} \Bigl[k^2q^2 + (k\cdot q)^2 \Bigr] g^{\mu \alpha}g^{\nu \beta}
-\frac{4(1-y)yz}{\Omega^2(1,y,z,1)} (k\cdot q) g^{\mu \alpha}k^{\nu}q^{\beta}\Biggr]
\Biggr\},
\end{split}
%\]
\eeq

\beq \label{lblcross}
\begin{split}
{\cal C}^{\alpha\beta\mu\nu}&=\int \frac{d^4p}{\pi^2i}C^{\alpha\beta\mu\nu}
=8\int_0^1 {dx}\int_0^1 {dy} \int_0^1 {dz}
\Biggl\{\biggl[\frac{x(1+x)}{\Omega(x,y,1,1)}
+\frac{x(1-x)}{\Omega(x,y,0,1)}
%\eeq
\\
%\[
&+ \frac{x(1-x)}{\Omega(x,1,z,1)}\biggr]
 g_{\mu \alpha}k_{\nu}q_{\beta}
-\frac{x^2y}{\Omega(x,y,z,1)}
(k\cdot q) g_{\mu \alpha}g_{\nu \beta}
+\frac{2x^2y}{\Omega^2(x,y,z,1)}
%\]
\\
%\[
&\times\Bigl[k^2(1-xy) + q^2(1-x+xyz) + (k\cdot q)(1-xy)(1-x+xyz)\Bigr]
 g_{\mu \alpha}k_{\nu}q_{\beta}\Biggr\},
\end{split}
%\]
\eeq

\beq
\Omega(x,y,z,\xi) = m^2 - k^2 xy(1-xy) - q^2x(1-yz)(1-x+xyz)
- 2k\cdot qxy[1-x-z(1-xy)]\xi,
\eeq

\noindent
and $g^{\mu\nu}=(1,-1,-1,-1)$.

The $\gamma$-matrix structures in \eq{lblskel} are antisymmetric in $(\alpha,\beta)$ and $(\mu,\nu)$, and we have thrown away all symmetric in $(\alpha,\beta)$ and $(\mu,\nu)$ terms in \eq{lblladd} and \eq{lblcross}. We have also combined terms that coincide after antisymmetrization, and deleted even in $k$ and $q$ terms that disappear anyway after substitution in the odd in these momenta integral in \eq{lblskel}. As a result we automatically subtracted symmetric in $k$ and $q$ logarithmically divergent contribution in $S^{\alpha\beta\mu\nu}$, and the result in \eq{gauginvblb} is finite and gauge invariant.

Next we substitute the light by light scattering tensor in \eq{lblskel},  and introduce two new Feynman parameters $t$ and $u$ to combine the upper photon propagators, the electron propagator, and the denominator $\Omega(x,y,z,\xi)$ in the integral representations in \eq{lblladd} and \eq{lblcross} of the light by light scattering tensor

\beq \label{combime}
(1-u)\left[(1-t)k^2+t(k^2-2mk_0)\right]+u
\left[\frac{\Omega(x,y,z,\xi)}{-xy(1-xy)}\right]
=\left(k-Q\right)^2-\Delta,
\eeq

\noindent
where $Q=qd+\tau$, $\Delta=g(-q^2+a^2)$, $\tau=m(1-u)t$, and

\beq
d=\xi u\left[z-\frac{1-x}{1-xy}\right],\quad
g=\frac{u(1-yz)(1-x+xyz)}{y(1-xy)}-d^2,
%\eeq
%\beq
\quad
a^2=\frac{1}{g}\left[\tau^2+\frac{m^2u}{xy(1-xy)}\right].
\eeq

\noindent
After the Wick rotation and integration over $k$ and $q$ (we return to dimensionless momenta here) we obtain an expression for the light by light contribution to HFS

\beq \label{twofirsttables}
\Delta E=2\Delta E_L+\Delta E_C\equiv\left(2\Delta \epsilon_L+\Delta \epsilon_C\right)\frac{\alpha^2(Z\alpha)}{\pi}E_F,
\eeq

\noindent
where

\beq
\Delta \epsilon_{L(C)}=\sum_i \Delta \epsilon_{L(C)}^{(i)}.
\eeq

\noindent
The integrals $\Delta \epsilon_{L(C)}^{(i)}$ arise in calculations of the ladder and crossed diagram contributions and have the general form

\beq  \label{intreprp}
\Delta \epsilon_{L(C)}^{(i)}=\int_0^\infty dq\int_0^1dx\int_0^1dy\int_0^1dz\int_0^1dt\int_0^1du\int_0^1d\xi {\cal J}_{L(C)}^{(i)}.
\eeq

\noindent
The integrands ${\cal J}_L^{(i)}$ and ${\cal J}_C^{(i)}$ are collected in Tables \ref{ladderlblqd} and \ref{crossblqd}, respectively. Notice that not all Feynman parameters arise in all integrands in Tables \ref{ladderlblqd} and \ref{crossblqd}. Some parameters just do not arise in particular integrals, or take a fixed value, for example, $x=1$ in the ladder light by light scattering diagram diagram, see \eq{lblladd}. As a result the expressions for $\Delta$ in the Tables are simpler than the general expression below \eq{combime}.

The third columns in Tables \ref{ladderlblqd} and \ref{crossblqd} contain separate integrals $\Delta\epsilon_L^{(i)}$ and $\Delta\epsilon_C^{(i)}$, and respective sums in the last lines. The sum of the ladder and crossed diagram contributions in \eq{twofirsttables} nicely reproduces the old result \cite{eks1991,kn199496}

\beq
\Delta E=-0.472~514~(1)\frac{\alpha^2(Z\alpha)}{\pi}E_F,
\eeq

\noindent
for light by light scattering contribution to HFS.

\begin{table}[htb]
\caption{First Set of Ladder Light by Light Integrals\footnote{
In this table
\[
d=\xi uz, \quad\tau=(1-u)t, \quad g=\frac{uz(1-yz)}{1-y}-d^2,
\quad a^2=\frac{1}{g}\left[\tau^2+\frac{u}{y(1-y)}\right],
\quad \Delta=g(q^2+a^2).
\]
}}
\begin{ruledtabular}
\begin{tabular}{lldd}
$i$ & ${\cal J}_{L}^{(i)}$ & \mbox{$\Delta \epsilon_L^{(i)}$} & \mbox{$\Delta \epsilon_L^{{vp}(i)}$}
\\
\colrule
$1$      & $-\frac{8}{\pi^2}
(1-t)(1-u)^2\left(\frac{1}{2\Delta}
+\frac{q^2d^2}{\Delta^2}\right)_{|z=1}$  & $-0.6255\ldots$  & $-0.914\ldots$
\\
$2$      &   $~~\frac{8}{\pi^2}
yz(1-t)(1-u)^2\left(\frac{1}{2\Delta}+ \frac{q^2d^2}{\Delta^2}\right)$
 & $0.1498\ldots$  &  $0.215\ldots$
\\
$3$      &   $~~\frac{4}{\pi^2}\frac{2-y+y^2z}{1-y}(1-t)(1-u)^2
\left(\frac{1}{\Delta}-\frac{\tau^2}{\Delta^2}\right)_{|\xi=1}$
 & $3.2252\ldots$  &  $6.403\ldots$
\\
$4$      &   $~~\frac{4}{\pi^2}
\frac{(1-u)d}{\Delta}_{|\xi=1}$
 & $0.0697\ldots$  & $0.085\ldots$
\\
$5$      &   $~~\frac{4}{\pi^2}q^2
\frac{yz}{1-y}\frac{(1-t)(1-u)^2d}{\Delta^2}_{|\xi=1}$
 & $0.1628\ldots$  &  $0.364\ldots$
\\
$6$      &   $~~\frac{4}{\pi^2}
\frac{2-y+yz}{1-y}(1-t)(1-u)^2\left(\frac{1}{2\Delta}+ \frac{q^2d^2}{\Delta^2}\right)_{|\xi=1}$
 & $2.0905\ldots$  &   $3.927\ldots$
\\
$7$      &   $-\frac{12}{\pi^2}
\frac{y(1-z)}{1-y}(1-t)(1-u)^2\left(\frac{3}{2\Delta}
+\frac{q^2d^2-\tau^2}{\Delta^2}\right)_{|\xi=1}$
 & $-3.8178\ldots$  &  $-8.070\ldots$
\\
$8$      &   $-\frac{8}{\pi^2}(1-z)u(1-u)\left(\frac{3}{2\Delta}
+\frac{q^2d^2-\tau^2}{\Delta^2}\right)_{|\xi=1}$
 & $-0.3303\ldots$  &  $-0.669\ldots$
\\
$9$      &   $-\frac{8q^2}{\pi^2}
\frac{y^2z^2(1-z)}{(1-y)^2}(1-t)u(1-u)^2\left(\frac{3}{2\Delta^2}
+\frac{2(q^2d^2-\tau^2)}{\Delta^3}\right)_{|\xi=1}$
 & $-0.4842\ldots$  &  $-1.078\ldots$
\\
$10$      &   $-\frac{8q^2}{\pi^2}
\frac{yz(1-z)}{1-y}\frac{u(1-u)d}{\Delta^2}_{|\xi=1}$
 & $-0.0193\ldots$  &  $-0.036\ldots$
\\
$11$      &   $-\frac{8q^2}{\pi^2}
\frac{yz(1-z)}{1-y}(1-t)u(1-u)^2d\left(\frac{3}{2\Delta^2} +  \frac{2q^2d^2}{\Delta^3}\right)_{|\xi=1}$
 & $-0.0105\ldots$  & $-0.019\ldots$
\\
$12$      &   $-\frac{8q^2}{\pi^2}
\frac{yz(1-z)}{1-y}(1-t)u(1-u)^2d\left(\frac{1}{\Delta^2} - \frac{2\tau^2}{\Delta^3}\right)_{|\xi=1}$
 & $-0.0058\ldots$ &  $-0.010\ldots$
\\
\colrule
$\sum_i$      &   & $0.4045\ldots$ &   $0.195\ldots$
\end{tabular}
\end{ruledtabular}
\label{ladderlblqd}
\end{table}

\begin{table}[htb]
\caption{First Set of Crossed Light by Light Integrals\footnote{
In this table
\[
d=u\left[z-\frac{1-x}{1-xy  }\right],\quad \tau=(1-u)t,
\quad g=\frac{u(1-yz)(1-x+xyz)}{y(1-y)}-d^2,
\]
\[
a^2=\frac{1}{g}\left[\tau^2+\frac{u}{xy(1-xy)}\right], \quad\Delta=g(q^2+a^2).
\]
}}
\begin{ruledtabular}
\begin{tabular}{lldd}
$i$ & ${\cal J}_{C}^{(i)}$ & \mbox{$\Delta \epsilon_C^{(i)}$} & \mbox{$\Delta \epsilon_C^{{vp}(i)}$}
\\
\colrule
$1$      & $-\frac{2}{\pi^2}\frac{1+x}{y(1-xy)}(1-t)(1-u)^2
\left(\frac{1}{\Delta} -\frac{\tau^2}{\Delta^2}\right)_{|z=1}$
& $-1.6733\ldots$  & $-3.294\ldots$
\\
$2$      & $-\frac{2}{\pi^2}\frac{1-x}{y(1-xy)}(1-t)(1-u)^2
\left(\frac{1}{\Delta} - \frac{\tau^2}{\Delta^2}\right)_{|z=0}$
& $-0.2729\ldots$ &  $-0.405\ldots$
\\
$3$      & $-\frac{2}{\pi^2}(1-t)(1-u)^2
\left(\frac{1}{\Delta} - \frac{\tau^2}{\Delta^2}\right)_{|y=1}$
&$-0.3665\ldots$ &  $-0.818\ldots$
\\
$4$      & $-\frac{2}{\pi^2}\frac{x}{1-xy}(1-t)(1-u)^2
\left(\frac{1}{\Delta} + \frac{2q^2d^2}{\Delta^2}\right)$
&$-0.2997\ldots$ &  $-0.431\ldots$
\\
$5$      & $-\frac{4}{\pi^2}\frac{u(1-u)}{y(1-xy)}
\left(-\frac{1}{\Delta} +\frac{\tau^2}{\Delta^2}\right)$
& $0.3460\ldots$ & $0.387\ldots$
\\
$6$      & $~~\frac{4}{\pi^2}\frac{1-x+xyz}{y(1-xy)^2}
(1-t)u(1-u)^2
\left(\frac{q^2}{\Delta^2} - \frac{2q^2\tau^2}{\Delta^3}\right)$
& $0.9869\ldots$ &  $2.943\ldots$
\\
$7$      & $~~\frac{4}{\pi^2}\frac{1-x+xyz}{y(1-xy)}
(1-t)u(1-u)^2d\left(\frac{q^2}{\Delta^2} - \frac{2q^2\tau^2}{\Delta^3}\right)$
& $-0.0020\ldots$ & $-0.004\ldots$
\\
\colrule
$\sum_i$    &   & $-1.2816\ldots$ & $-1.623\ldots$
\end{tabular}
\end{ruledtabular}
\label{crossblqd}
\end{table}

Let us calculate contributions to HFS generated by the diagrams with polarization insertions in Figs. \ref{downphot} and \ref{upphot}.  We use the well known integral representation for the polarization operator (see, e.g., \cite{egs2007})

\beq \label{ploop}
\Pi(q^2)=
\frac{\alpha}{\pi }\int_0^1dvv^2\left(1-\frac{v^2}{3}\right)
\frac{q^2}{q^2(1-v^2)+4},
\eeq

\noindent
where the dimensionless momentum $q$ is Euclidean.

Momentum $q$ in the integrands in Tables \ref{ladderlblqd} and \ref{crossblqd} is also Euclidean and to account for the polarization operator insertions in both lower photon lines in Fig. \ref{downphot} it is sufficient to insert the factor $2\Pi(q^2)$ in the integrands in Tables \ref{ladderlblqd} and \ref{crossblqd}. Similarly to \eq{twofirsttables} respective contributions to HFS can be written as

\beq \label{downvpins}
\Delta E_d=2\Delta E^{vp}_L+\Delta E^{vp}_C\equiv\left(2\Delta \epsilon^{vp}_L+\Delta \epsilon^{vp}_C\right)\frac{\alpha^3(Z\alpha)}{\pi^2}E_F,
\eeq

\noindent
where

\beq
\Delta \epsilon^{vp}_{L(C)}=\sum_i \Delta \epsilon_{L(C)}^{{vp}(i)}.
\eeq

\noindent
The fourth columns in Tables \ref{ladderlblqd} and \ref{crossblqd} contain the integrals $\Delta\epsilon_L^{{vp}(i)}$ and $\Delta\epsilon_C^{{vp}(i)}$, and their sums in the last row. Collecting these contributions we obtain the total contribution to HFS generated by the diagrams in Fig. \ref{downphot}

\beq \label{vpdown}
\Delta E_d=-1.2326(5)\frac{\alpha^3(Z\alpha)}{\pi^2}E_F.
\eeq

Calculation of the contributions to HFS generated by the diagrams in Fig. \ref{upphot} follows the same general route as for the diagrams with polarization insertions in the lower photons. We again parameterize contribution to HFS generated  the light by light scattering diagrams  in Fig. \ref{lbl} exactly like in \eq{twofirsttables}. However, now it is convenient first to integrate analytically over momentum $q$ in the integrals in \eq{intreprp}. As a result the separate contributions to HFS acquire the form

\beq  \label{intreprpwq}
\Delta \epsilon_{L(C)}^{(i)}=
\int_0^1dy\int_0^1dz\int_0^1dt\int_0^1du\int_0^1d\xi {\cal K}_{L(C)}^{(i)}.
\eeq

\noindent
The integrands ${\cal K}_L^{(i)}$ and ${\cal K}_C^{(i)}$ are collected in Tables \ref{ladderlblqup} and \ref{crossblqup}.

\begin{table}[htb]
\caption{Second Set of Ladder Light by Light Integrals\footnote{
In this table
\[
d=\xi uz, \quad\tau=(1-u)t,\quad g=\frac{uz(1-yz)}{1-y}-d^2,\quad a^2=\frac{1}{g}\left[\tau^2+\frac{u}{y(1-y)}\right].
\]
}}
\begin{ruledtabular}
\begin{tabular}{lldd}
$i$ & ${\cal K}_{L}^{(i)}$ & \mbox{$\Delta \epsilon_L^{(i)}$} & \mbox{$\Delta \epsilon_L^{{vp}(i)}$}
\\
\colrule
$1$      & $-\frac{2}{\pi}(1-t)(1-u)^2\left(\frac{1}{ag}
+\frac{d^2}{ag^2}\right)_{|z=1}$
& $-0.6255\ldots$ &  $-0.4631\ldots$
\\
$2$      &   $~~\frac{2}{\pi}
yz(1-t)(1-u)^2\left(\frac{1}{ag}+ \frac{d^2}{ag^2}\right)$
& $0.1498\ldots$  &  $0.1044\ldots$
\\
$3$      &   $~~\frac{1}{\pi}\frac{2-y+y^2z}{1-y}(1-t)(1-u)^2
\left(\frac{2}{ag}-\frac{\tau^2}{a^3g^2}\right)_{|\xi=1}$
& $3.2252\ldots$ &  $2.9251\ldots$
\\
$4$      &   $~~\frac{2}{\pi}
\frac{(1-u)d}{ag}_{|\xi=1}$
& $0.0697\ldots$ & $0.1083\ldots$
\\
$5$      &   $~~\frac{1}{\pi}
\frac{yz}{1-y}\frac{(1-t)(1-u)^2d}{ag^2}_{|\xi=1}$
& $0.1628\ldots$ & $0.1273\ldots$
\\
$6$      &   $~~\frac{1}{\pi}
\frac{2-y+yz}{1-y}(1-t)(1-u)^2\left(\frac{1}{ag}+ \frac{d^2}{ag^2}\right)_{|\xi=1}$
& $2.0905\ldots$ &  $1.6319\ldots$
\\
$7$      &   $-\frac{3}{\pi}
\frac{y(1-z)}{1-y}(1-t)(1-u)^2\left(\frac{3}{ag}
+\frac{d^2}{ag^2}-\frac{\tau^2}{a^3g^2}\right)_{|\xi=1}$
&$-3.8178\ldots$  &  $-3.3474\ldots$
\\
$8$      &   $-\frac{2}{\pi}(1-z)u(1-u)\left(\frac{3}{ag}
+\frac{d^2}{ag^2}-\frac{\tau^2}{a^3g^2}\right)_{|\xi=1}$
& $-0.3303\ldots$ & $-0.5174\ldots$
\\
$9$      &   $-\frac{1}{\pi}
\frac{y^2z^2(1-z)}{(1-y)^2}(1-t)u(1-u)^2\left(\frac{3}{ag^2}
+\frac{3d^2}{ag^3}-\frac{\tau^2}{a^3g^3}\right)_{|\xi=1}$
& $-0.4842\ldots$ &  $-0.4946\ldots$
\\
$10$      &   $-\frac{2}{\pi}
\frac{yz(1-z)}{1-y}\frac{u(1-u)d}{ag^2}_{|\xi=1}$
& $-0.0193\ldots$ & $-0.0307\ldots$
\\
$11$      &   $-\frac{3}{\pi}
\frac{yz(1-z)}{1-y}(1-t)u(1-u)^2d\left(\frac{1}{ag^2} +  \frac{d^2}{ag^3}\right)_{|\xi=1}$
& $-0.0105\ldots$ &  $-0.0168\ldots$
\\
$12$      &   $-\frac{1}{\pi}
\frac{yz(1-z)}{1-y}(1-t)u(1-u)^2d\left(\frac{2}{ag^2} - \frac{\tau^2}{a^3g^3}\right)_{|\xi=1}$
& $-0.0058\ldots$ &  $-0.0092\ldots$
\\
\colrule
$\sum_i$      &   $$ & $0.4045\ldots$ &  $0.0177\ldots$
\end{tabular}
\end{ruledtabular}
\label{ladderlblqup}
\end{table}

\begin{table}[htb]
\caption{Second Set of Crossed Light by Light Integrals\footnote{
In this table
\[
d=u\left[z-\frac{1-x}{1-xy}\right],\; \tau=(1-u)t,\; g=\frac{u(1-yz)(1-x+xyz)}{y(1-y)}-d^2,\; a^2=\frac{1}{g}\left[\tau^2+\frac{u}{xy(1-xy)}\right].
\]
}}
\begin{ruledtabular}
\begin{tabular}{lldd}
$i$ & ${\cal K}_{C}^{(i)}$ & \mbox{$\Delta \epsilon_C^{(i)}$} & \mbox{$\Delta \epsilon_C^{{vp}(i)}$}
\\
\colrule
$1$      & $-\frac{1}{2\pi}\frac{1+x}{y(1-xy)}(1-t)(1-u)^2
\left(\frac{2}{ag} -\frac{\tau^2}{a^3g^2}\right)_{|z=1}$
& $-1.6733\ldots$  & $-1.5131\ldots$
\\
$2$      & $-\frac{1}{2\pi}\frac{1-x}{y(1-xy)}(1-t)(1-u)^2
\left(\frac{2}{ag} - \frac{\tau^2}{a^3g^2}\right)_{|z=0}$
& $-0.2729\ldots$  & $-0.2646\ldots$
\\
$3$      & $-\frac{1}{2\pi}(1-t)(1-u)^2
\left(\frac{2}{ag} - \frac{\tau^2}{a^3g^2}\right)_{|y=1}$
& $-0.3665\ldots$  &  $-0.3084\ldots$
\\
$4$      & $-\frac{1}{\pi}\frac{x}{1-xy}(1-t)(1-u)^2
\left(\frac{1}{ag} + \frac{d^2}{ag^2}\right)$
& $-0.2997\ldots$  &  $-0.2088\ldots$
\\
$5$      & $~~\frac{1}{\pi}\frac{u(1-u)}{y(1-xy)}
\left(\frac{2}{ag} -\frac{\tau^2}{a^3g^2}\right)$
& $0.3460\ldots$  &  $0.5995\ldots$
\\
$6$      & $~~\frac{1}{2\pi}\frac{1-x+xyz}{y(1-xy)^2}
(1-t)u(1-u)^2
\left(\frac{2}{ag^2} - \frac{\tau^2}{a^3g^3}\right)$
& $0.9869\ldots$  & $0.8396\ldots$
\\
$7$      & $~~\frac{1}{2\pi}\frac{1-x+xyz}{y(1-xy)}
(1-t)u(1-u)^2d\left(\frac{2}{ag^2} - \frac{\tau^2}{a^3g^3}\right)$
& $-0.0020\ldots$  &  $-0.0038\ldots$
\\
\colrule
$\sum_i$      &   $$ & $-1.2816\ldots$ & $-0.8597\ldots$

\end{tabular}
\end{ruledtabular}
\label{crossblqup}
\end{table}

Insertion of the polarization operator in \eq{ploop} (with $q^2\to-k^2$) in the upper photon lines of the diagrams in Fig. \ref{upphot} is described by insertion inside the integrands in Tables \ref{ladderlblqup} and \ref{crossblqup} of the factor

\beq
2\left(\frac{\alpha}{\pi}\right)\int_0^1 dw\int_0^1dv\frac{v^2}{1-v^2}\left(1-\frac{v^2}{3}\right),
\eeq

\noindent
introduction of an additional Feynman parameter $w$ and the substitution

\beq
a^2\to a^2(w)=a^2+\frac{4w(1-t)(1-u)}{g(1-v^2)}.
\eeq

The extra Feynman parameter $w$ does not arise (it is effectively equal one) in the vacuum polarization integrals in the $4$th, $8$th, and $10$th rows in Table \ref{ladderlblqup}, and in the $5$th row in Table \ref{crossblqup}. For these integrals the substitution reduces to  $a^2\to a^2(1)$.

Similarly to \eq{downvpins} we represent the contributions to HFS generated by the diagrams in Fig. \ref{upphot} in the form

\beq
\Delta E_u=2\Delta E^{vp}_L+\Delta E^{vp}_C\equiv\left(2\Delta \epsilon^{vp}_L+\Delta \epsilon^{vp}_C\right)\frac{\alpha^3(Z\alpha)}{\pi^2}E_F,
\eeq

\noindent
where

\beq
\Delta \epsilon^{vp}_{L(C)}=\sum_i \Delta \epsilon_{L(C)}^{{vp}(i)}.
\eeq

\noindent
The fourth columns in Tables \ref{ladderlblqup} and \ref{crossblqup} contain the integrals $\Delta\epsilon_L^{{vp}(i)}$ and $\Delta\epsilon_C^{{vp}(i)}$, and their sums in the last line. Collecting these contributions we obtain the total contribution to HFS generated by the diagrams in Fig. \ref{upphot}

\beq \label{vpup}
\Delta E_u=-0.8242(1)\frac{\alpha^3(Z\alpha)}{\pi^2}E_F.
\eeq

\section{Conclusions}

Collecting results in \eq{vpdown} and \eq{vpup} we obtain the total contribution to HFS generated by the polarization insertions in Figs. \ref{downphot} and \ref{upphot}

\beq \label{vptot}
\Delta E_{vp}=-2.056(1)\frac{\alpha^3(Z\alpha)}{\pi^2}E_F,
\eeq

\noindent
or numerically

\beq
\Delta E_{vp}=-2.63~\text{Hz}.
\eeq

Three-loop contribution to HFS containing closed electron loops and factorized one-loop radiative insertions in the electron line were calculated earlier \cite{es2003,es2004,es2007}. Combining those corrections with the result in \eq{vptot} we obtain the sum of all gauge invariant three-loop radiative corrections to HFS calculated thus far

\beq
\Delta E_t=-3.338(1)\frac{\alpha^3(Z\alpha)}{\pi^2}E_F,
\eeq

\noindent
or numerically

\beq
\Delta E_t=-4.28~\text{Hz}.
\eeq

\noindent
Work on calculation of the remaining three-loop contributions to HFS is now is progress.

\begin{acknowledgments}

This work was supported by the NSF grant PHY-1066054. The work of V. S. was also supported in part by the RFBR grant 12-02-00313 and by the DFG grant GZ: HA 1457/7-2.

\end{acknowledgments}

\end{document}